\documentclass[showpacs,preprintnumbers,amsmath,amssymb]{revtex4}
\usepackage{amsmath}
\usepackage{graphicx}
\usepackage{subfig}
\usepackage{hyperref}
\usepackage{amssymb}
\usepackage[english]{babel}
\usepackage{epsfig}
\usepackage{wasysym}

\def\nn{\nonumber \\}

\newcommand{\be}{\begin{equation}}

\newcommand{\ee}{\end{equation}}

\newcommand{\bea}{\begin{eqnarray}}

\newcommand{\eea}{\end{eqnarray}}

\newcommand{\beq}{\begin{eqnarray}}

\newcommand{\eeq}{\end{eqnarray}}

\begin{document}
\title{Dynamics of scalar perturbations in $f(R, T)$ gravity \\}
\author{ F. G. Alvarenga$^{(a)}$, A. de la Cruz-Dombriz$^{(b,c)}$, M. J. S. Houndjo$^{(a,d)}$,  
M. E. Rodrigues$^{(e)}$, and
D. S\'{a}ez-G\'{o}mez$^{(b,c,f)}$}
\affiliation{$^{a}$ \, Departamento de Engenharia e Ci\^{e}ncias Naturais - CEUNES -
Universidade Federal do Esp\'irito Santo -  CEP 29933-415 - S\~ao Mateus/ ES, Brazil.\\
$^b$\ Astrophysics, Cosmology and Gravity Centre (ACGC),
University of Cape Town, Rondebosch 7701, Cape Town, South Africa.\\
$^c$\ Department of Mathematics and Applied Mathematics, University of Cape Town, Rondebosch 7701, Cape Town, South Africa .\\
$^d$ \, Institut de Math\'{e}matiques et de Sciences Physiques (IMSP) 01 BP 613 Porto-Novo, B\'{e}nin.\\
$^{e}$\, Universidade Federal do Esp\'irito Santo - Centro de Ci\^{e}ncias Exatas - Departamento de F\'isica\\
Av. Fernando Ferrari s/n - Campus de Goiabeiras CEP29075-910 - Vit\'oria/ES, Brazil.\\
$^f$\ Fisika Teorikoaren eta Zientziaren Historia Saila, Zientzia eta Teknologia Fakultatea,\\
Euskal Herriko Unibertsitatea, 644 Posta Kutxatila, 48080 Bilbao, Spain, EU}
\begin{abstract}
In the context of $f(R,T)$ theories of gravity, we study the evolution of scalar cosmological perturbations in the metric formalism. 
According to restrictions on the background evolution, a specific model within these theories is assumed in order to guarantee the 
standard continuity equation. Using a completely general procedure, we find the complete set of differential equations for the matter density perturbations. In the case of sub-Hubble modes, the density contrast evolution reduces to a second-order equation. 
We show that for well-motivated 
$f(R, T)$ Lagrangians the quasi-static approximation yields to very different results from the ones 
derived in the frame of the Concordance $\Lambda$CDM model constraining severely the viability of 
such theories. 
\end{abstract}
\pacs{ 98.80.-k, 04.50.Kd, 95.36.+x}
\maketitle
\section{Introduction}
It is a well-known fact that modifying the law of gravity renders possible explanations for the acceleration mechanism of the Universe \cite{0601213}-\cite{3-6antonio2011}. However it is far from clear which class of dark energy (DE) theories will finally prevail and all the viable mechanism must be studied very carefully. Whereas the theories explaining the accelerated expansion of the Universe in the framework of the general relativity (GR) \cite{2gannouji} are usually dubbed as DE models, theories in the framework of modified gravity are more specifically referred to as modified gravity DE theories. 
\par
\vspace{0.2cm}

In this letter we consider a class of modified gravity theories in which the gravitational action contains a general function $f(R,T)$, where $R$ and $T$ denote the Ricci scalar and the trace of the energy-momentum tensor, respectively. This kind of modified gravity was introduced first in \cite{fRTpaper} 
where some significant results were obtained. In the framework of $f(R,T)$ gravity, some cosmological aspects have been already explored: the reconstruction of cosmological solutions, where late-time acceleration is accomplished was studied in Ref.~\cite{stephaneseul3} and the energy conditions analyzed in Ref.~\cite{flavio}. The thermodynamics of Friedmann-Lema\^itre-Robertson-Walker (FLRW) spacetimes has been studied in Ref.~\cite{thermo1}, and also the possibility of the occurrence of future singularities  (see Ref.~\cite{juliano}).
So far, a serious shortcoming in this kind of theory has been the non-conservation of the energy-momentum tensor. In this paper we circumvent this problem
by showing  that 
functions $f(R,T)$ can always be constructed to be consistent with the energy-momentum tensor standard conservation. In the following we shall assume separable algebraic functions of the form $f(R,T)=f_1(R)+f_2(T)$. Within this special choice, the function $f_2(T)$ is obtained by imposing the conservation of the energy-momentum tensor. 
\par
\vspace{0.2cm}

Once the cosmic background evolution is known, the following step consists of determining the evolution of matter cosmological perturbations. 
The analysis of perturbed field equations by decomposing linear perturbations in scalar, vector and tensor modes, led to a better understanding of the stability and features of the Robertson-Walker spacetime and proved to be a required tool to analise the density contrast growth and the integrated Salch-Wolfe effect. Second (and higher) order perturbative terms with respect to the background are usually considered negligible since the perturbations are assumed to be small in order to preserve the global homogeneity and isotropy of Robertson-Walker geometry and therefore the linear terms are generally enough to encapsulate the small departure from the background. Seminal references \cite{Bardeen} focused their attention to GR and found that the evolution of the density contrast obeyed a second order differential equation which in the subHubble limit is scale-independent. In such regime, the density contrast grows as the cosmological scale factor for early times whereas at late times requires to be fitted numerically \cite{Linder}. 

\par
\vspace{0.2cm}
Nonetheless the growth of structures is manifestly dependent on the gravitational theory under consideration. This fact can be used to test alternative theories of gravity in order to find out whether those theories are in agreement with GR standard predictions \cite{3-gannouji} and experimental data \cite{other_collab} and thus curing the so-called {\it degeneracy problem}  \cite{Alvaro_ERE2012} that some modified gravity theories suffer at the background level.  
This sort of work has been developed in the last years but mainly for the $f(R)$ gravity scenarios  \cite{otras,  fR_perturbations}. 
In this realm, the evolution of scalar cosmological perturbation for $f(R)$ theories in the metric formalism 
proved that $f(R)$ theories mimicking the standard cosmological expansion, usually provide a different matter power spectrum from that predicted by the $\Lambda$CDM model  \cite{Dombriz_PRD2008}. Still in the framework of $f(R)$ theories of gravity, the growth of matter perturbations at low redshifts was shown to be different from that of scalar-tensor theories \cite{gannouji}. For further details about cosmological perturbations within $f(R)$ gravity see \cite{restantf(R)}.
Therefore the dynamics of cosmological scalar perturbations is a powerful tool to constrain the viability of
the pleiad of  
modified gravity theories in literature, by comparing their density contrast evolution with GR expected features \cite{4-gannouji, Comment,Dombriz2013}. 
\par
\vspace{0.2cm}

Nonetheless, no full attention has been yet paid to study the density contrast evolution in $f(R,T)$ theories. Extensive analysis have been  carried out in the framework of non-standard couplings between the geometry and the matter Lagrangian (see \cite{Nesseris:2008mq}).
For our purpose in this communication, the dynamics of linear perturbations are performed studying the problem of obtaining the exact equations for the evolution of matter density perturbations for $f_1(R)+f_2(T)$ type gravitational Lagrangians. More precisely, we shall assume for simplicity the algebraic function $f_1(R)$ to be the Einstein-Hilbert term $R$ and the trace dependent function $f_2(T)$ the one for which the covariant conservation of the energy-momentum is accomplished. 
When interested in sub-Hubble modes, the usual approach consists of studying 
the so-called quasi-static approximation where time derivatives of Bardeen's potentials are neglected, and only time derivatives involving density perturbations are kept \cite{Dombriz_PRD2008,restantf(R)}. Let us point out that this approximation may remove essential information about the evolution of the first-order perturbed fields \cite{Silvestri,Starobinsky} and therefore requires careful study when considered.
\par
\vspace{0.2cm}

The paper is organized as follows: 
in Section \ref{Section2}, we briefly review the state-of-the-art of $f(R,T)$ gravity. 
Section  \ref{Section3} is devoted to introduce the background cosmological equations for  $f(R,T)=f_1(R)+f_2(T)$ models as well as the condition to guarantee standard energy-momentum conservation for such models.
Then, Section \ref{Section4} addressed the calculation of the scalar perturbed equations for $f(R,T)=f_1(R)+f_2(T)$ models
while Section \ref{Section5} deals with the study of the quasi-static approximation for this kind of models. In Section \ref{Section6} we apply our
results to two particular models and numerical results are obtained and compared with the $\Lambda$CDM model. 
Finally in Section \ref{Section7} we conclude with the main 
conclusions of this investigation. 

\section{$f(R, T)$ gravity theories} \label{Section2}

Let us start by writing the general action for $f(R,T)$ gravities \cite{fRTpaper}, 
\be
S\,=\,S_G+S_m\,=\,\frac{1}{2\kappa^2}\int {\rm d}^4x\sqrt{-g}\ \left(f(R,T)+
\mathcal{L}_m\right)\ ,
\label{I.1}
\ee
where, $\kappa^2=8\pi G$, $R$ is the Ricci scalar and $T$ represents the trace of the energy-momentum tensor, i.e., $T=T^{\mu}_{\;\;\mu}$, while $\mathcal{L}_m$ is the matter Lagrangian. As usual the energy-momentum tensor is defined as,
\be
T_{\mu\nu}=\frac{2}{\sqrt{-g}}\frac{\delta S_m}{\delta g^{\mu\nu}}\,.
\label{I.2}
\ee
Then, by varying the action with respect to the metric field $g^{\mu\nu}$, the field equations are obtained,
\be
f_{R}(R,T)R_{\mu\nu}-\frac{1}{2}f(R,T)g_{\mu\nu}-\left(g_{\mu\nu}\Box-\nabla_{\mu}\nabla_{\nu}\right)f_{R}(R,T)=-\left(\kappa^2+f_T(R,T)\right)T_{\mu\nu}-f_{T}(R,T)\Theta_{\mu\nu}\ ,
\label{I.3}
\ee
where the subscripts on the function $f(R,T)$ mean differentiation with respect to $R$ or $T$, and the tensor  $\Theta_{\mu\nu}$ is defined as,
\be
\Theta_{\mu\nu}\,\equiv\, g^{\alpha\beta}\frac{\delta T_{\alpha\beta}}{\delta g^{\mu\nu}}=-2T_{\mu\nu}-g_{\mu\nu}\mathcal{L}_m+2g^{\alpha\beta}\frac{\delta\mathcal{L}_m}{\delta g_{\mu\nu}g_{\alpha\beta}}\ .
\label{I.4}
\ee
Note that for a regular $f(R,T)$ function, in absence of any kind of matter, the corresponding $f(R)$ gravity equations are recovered, and consequently the corresponding properties and the well-known solutions for $f(R)$ gravity are also satisfied by $f(R,T)$ theories in classical vacuum (for a review on $f(R)$ theories, see \cite{0601213}). Moreover, here we are interested to study the behavior of this kind of theories for spatially flat FLRW spacetimes, which  are expressed in comoving  coordinates by the line element,
\be
{\rm d}s^2\,=\,a^{2}(\eta)\left({\rm d}\eta^2-{\rm d}{\bf x}^2\right)
\label{I.5}
\ee
where $a(\eta)$ is the scale factor in conformal time $\eta$. Then, the main issue arises on the content of the Universe is given by through the energy-momentum tensor, defined in (\ref{I.2}). Since  we are interested on flat FLRW cosmologies, the usual content of the Universe (pressureless matter, radiation,... ) can be well described by perfect fluids, whose energy-momentum tensors take the form,
\be
T_{\mu\nu}=(\rho+p)u_{\mu}u_{\nu}-pg_{\mu\nu}\ .
\label{I.6}
\ee
Here $\rho$ and $p$ are the energy and pressure densities respectively, and $u^{\mu}$ is the four-velocity of the fluid, which satisfies $u_{\mu}u^{\mu}=1$, and in comoving coordinates is given by $u^{\mu}=(1,\ 0,\ 0,\ 0)$. Since $\mathcal{L}_m=p$, according to the definition suggested in Ref.~\cite{fRTpaper}, the tensor (\ref{I.4}) yields,
 \be
\Theta_{\mu\nu}= -2 T_{\mu\nu}-p\ g_{\mu\nu}\,\,.
\label{I.7}
\ee
Thus the equations motion become
\begin{eqnarray}
f_RR_{\mu\nu}-\frac{1}{2}fg_{\mu\nu}-\left(g_{\mu\nu}\Box-\nabla_{\mu}\nabla_{\nu}\right)f_R=-\left(\kappa^2-f_T\right)T_{\mu\nu}+f_Tpg_{\mu\nu}\,\,. \label{manu7}
\label{I.8}
\end{eqnarray}
where we have dropped the explicit dependences of $f$ in $R$ and $T$.
\par
It is straightforward to see that the usual continuity equation is not satisfied for the field equations \eqref{I.8}, and consequently the covariant derivative of the energy-momentum tensor is not null in general $\nabla_{\mu}T^{\mu\nu}\neq0$. In order to obtain the modified continuity equation, let us take the covariant derivative of the equation \eqref{I.8},
\[
\nabla^{\mu}\left[f_{R}R_{\mu\nu}-\frac{1}{2}f g_{\mu\nu}-\left(g_{\mu\nu}\Box-\nabla_{\mu}\nabla_{\nu}\right)f_{R}=-\left(\kappa^2+f_T\right)T_{\mu\nu}-f_{T}\Theta_{\mu\nu}\right]
\]
\[
\rightarrow f_{R}\nabla^{\mu}R_{\mu\nu}+R_{\mu\nu}\nabla^{\mu}f_{R}-\frac{1}{2}g_{\mu\nu}(f_{R}\nabla^{\mu}R+f_{T}\nabla^{\mu}T)-\left(g_{\mu\nu}\nabla^{\mu}\Box-\nabla^{\mu}\nabla_{\mu}\nabla_{\nu}\right)f_{R}=
\]
\be
\nabla^{\mu}\left[-\left(\kappa^2+f_T\right)T_{\mu\nu}-f_{T}\Theta_{\mu\nu}\right]\ .
\label{I.10}
\ee
Thus, using the identities $\nabla^{\mu}\left(R_{\mu\nu}-\frac{1}{2}Rg_{\mu\nu}\right)=0$, and $\left(\nabla_{\nu}\Box-\Box\nabla_{\nu}\right)f_{R}(R,T)=R_{\mu\nu}\nabla^{\mu}f_R$, the covariant derivative of the energy-momentum tensor needs to satisfy,
\be
\nabla^{\mu}T_{\mu\nu}=\frac{f_T}{\kappa^2+f_T}\left[\frac{1}{2}g_{\mu\nu}\nabla^{\mu}T-(T_{\mu\nu}+\Theta_{\mu\nu})\nabla^{\mu}\ln f_T-\nabla^{\mu}\Theta_{\mu\nu}\right]\,\,.
\label{I.11}
\ee
Hence, for a perfect fluid with an equation of state $p=w\rho$, being $w$ a constant, the $0-$component of the covariant derivative \eqref{I.11} turns out to become,
\be
\left[\kappa^2+\frac{w-3}{2}f_{T}-(1+w)T f_{TT}\right]\dot{T}+3(1+w)\left[H(\kappa^2-f_T)-2f_{TR}(4H\dot{H}+\ddot{H})\right]T=0\ .
\label{I.12}
\ee
where let us remind that $T=T^{\mu}_{\;\mu}=\rho-3p$. The last  equation differs from the usual continuity equation on the non-null right hand side (r.h.s.). Thus, it may lead to violations of the usual evolution 
of the different species in the Universe. Nevertheless,  in the next section 
we focus our attention on a model that keeps  the usual continuity equation unchanged.

\section{$f_1(R)+f_2(T)$ type theories}\label{Section3}

In this section, we choose the algebraic function $f(R,T)$ to be a sum of two independent functions 
\begin{eqnarray}
f(R,T)\,=\,f_1(R) + f_2(T)
\label{separable_function}
\end{eqnarray}
where $f_1(R)$ and $f_2(T)$, respectively depend on the curvature $R$ and the trace $T$.  The generalized Einstein equations from (\ref{manu7}) yield
\begin{eqnarray}
-3\mathcal{H}f^{'}_{1R_0}+3\mathcal{H}'f_{1R_0}-\frac{a^2}{2}f_{10}&=&-\kappa^2 a^2\rho_0+(1+c_s^2)\rho_0a^2f_{2T_0}+\frac{a^2}{2}f_{20}\,\,,\label{manu13}\\
f_{1R_0}^{''}+\mathcal{H}f^{'}_{1R_0}-(\mathcal{H}'+2\mathcal{H}^2)f_{1R_0}+\frac{a^2}{2}f_{10}&=&-\kappa^2 a^2c_s^2\rho_0-\frac{a^2}{2}f_{20}\label{manu14}\,\,\,.
\label{FLRWf2}
\,\,\,
\end{eqnarray}
where the prime holds for the derivative with respect to $\eta$, $\mathcal{H}\equiv a'/a$ and 
the subscript $0$ holds for unperturbed background quantities: 
$R_0$ denotes the scalar curvature corresponding to the unperturbed metric, 
$\rho_0$ the unperturbed energy density, with $f_{10}\equiv f_1(R_0)$, $f_{1R_0}\equiv {\rm d}f_1(R_0)/{\rm d}R_0$, $f_{20}\equiv f_2(T_0)$, $f_{2T_0}\equiv {\rm d}f_2(T_0)/{\rm d}T_0$ and $c_s^2=p_0/\rho_0$. The continuity equation (\ref{I.12}) for Lagrangians given by (\ref{separable_function}) yields
\begin{eqnarray}
\nabla_{\mu}T^{\mu}_{0\,\nu}=\frac{1}{\kappa^2-f_{2T_0}}\left[\delta^{\mu}_{\nu}\partial_{\mu}\left(\frac{1}{2}f_{20}+c_s^2\rho_0
f_{2T_0}\right)
+T_{0\,\nu}^{\;\mu}\partial_{\mu}f_{2T_0}\right]\,\,,
\label{manu14}
\end{eqnarray}
showing explicitly that the energy-momentum tensor is not a priori covariantly conserved in $f(R,T)$ theories. Thus, for these theories, the test particles moving in a gravitational field do not follow geodesic lines. By exploring  the equation (\ref{manu14}) for $\nu=0$ component, one gets 
\begin{eqnarray}
\rho^{'}_{0}+
3\mathcal{H}\rho_0(1+c_s^2)=\frac{1}{\kappa^2-f_{2T_0}}\left[ (1+c_s^2)\rho_0f'_{2T_0}+c_s^2\rho'_0f_{2T_0}+\frac{1}{2}f'_{20}\right]      \,\,.
\end{eqnarray}
Note that whether $f_2$ vanishes (i.e., $f(R)$ theories) or characterizes a non-running cosmological constant, both $f^{'}_{2T_0}$ and $f_{2T_0}$ vanish, and then the continuity equation 
in these scenarios becomes
\begin{eqnarray}\label{eqconserv}
\rho_0^{'}+3\mathcal{H}\left(1+c_s^2\right)\rho_0=0\,\,\,.
\end{eqnarray}

In order to Lagrangians such as (\ref{separable_function}) consistent with the standard conservation equation, the r.h.s. of (\ref{manu14}) has to vanish leading to 
the differential equation
\begin{eqnarray}
\left(1+c_s^2\right)T_0f_{2T_0T_0}+\frac{1}{2}\left(1-c_s^2\right)f_{2T_0}=0\,\,,
\end{eqnarray}
where $c_s^2\neq 1/3$. The general solution of this differential equation reads
\begin{eqnarray}
f_2(T_0)=\alpha T_0^{\frac{1+3c_s^2}{2\left(1+c_s^2\right)}}+\beta\,\,,
\label{f2T}
\end{eqnarray}
where $\alpha$ and $\beta$ are integration constants. In the case of a barotropic equation of state $c_s^2=0$, i.e., dust, the model (\ref{f2T}) becomes 
\begin{eqnarray}
f_2(T_0)=\alpha T_0^{1/2}+\beta
\label{viable_model_dust}
\end{eqnarray}
This function represents the unique Lagrangian that satisfies the usual continuity equation (\ref{eqconserv}) within the class of models given by expression 
(\ref{separable_function}).

\section{Perturbations in $f(R,T)$ theories}\label{Section4}

Let us consider the scalar perturbations of a flat FLRW metric in the longitudinal gauge
\begin{eqnarray}
{\rm d}s^2=a^2(\eta)\left[(1+2\Phi){\rm d}\eta^2-(1-2\Psi){\rm d}{\bf x}^2\right]\,\,,
\label{manu17}
\end{eqnarray}
where $\Phi \equiv \Phi(\eta,{\bf x})$ and $\Psi \equiv \Psi(\eta,{\bf x})$ are the scalar perturbations. The components of perturbed energy-momentum tensor in this gauge are given by
\begin{eqnarray}
\hat{\delta} T^{0}_{\;0}= \hat{\delta}\rho = \rho_0 \delta \,,\quad \hat{\delta} T^{i}_{\;j}=-\hat{\delta} p\;\delta^{i}_{\;j}=-c^2_s\rho_0\delta^{i}_{\;j}\delta\,,\quad \hat{\delta} T^{0}_{\;i}=-\hat{\delta} T^{i}_{\;0}=-\left(1+c_s^2\right)\rho_0\partial_iv\,, \label{manu18}
\end{eqnarray}
where $v$ denotes the potential for the velocity perturbations. Using the model (\ref{separable_function}), the perturbed metric (\ref{manu17}) and the perturbed energy-momentum tensor (\ref{manu18}), the first order perturbed equations  reads
\begin{eqnarray}
f_{1R_0}\hat{\delta} G^{\mu}_{\nu}+\left(R^{\;\mu}_{0\nu}+\nabla^{\mu}\nabla_{\nu}-\delta^{\mu}_{\nu}\Box\right)f_{1R_0R_0}\hat{\delta} R+\left[\left(\hat{\delta} g^{\mu\alpha}\right)\nabla_{\nu}\nabla_{\alpha}-\delta^{\mu}_{\nu}\left(\hat{\delta} g^{\alpha\beta}\right)\nabla_{\alpha}\nabla_{\beta}\right]f_{1R_0}
\nonumber\\
-\left[g_{0}^{\alpha\mu}\left(\hat{\delta}\Gamma^{\gamma}_{\alpha\nu}\right)-
\delta^{\mu}_{\nu}g_{0}^{\alpha\beta}
\left(\hat{\delta}\Gamma^{\gamma}_{\beta\alpha}\right)\right]\partial_{\gamma}
f_{1R_0}=-\left(\kappa-f_{2T_0}\right)\hat{\delta} T^{\mu}_{\nu}\nonumber\\ 
+\left[\frac{1}{2}(1-c_{s}^2)f_{2T_{0}}\delta^{\mu}_{\nu}
+(1-3c_{s}^2)(1+c_{s}^2)\rho_{0}f_{2T_{0}T_{0}}u^{\mu}u_{\nu}
\right]\hat{\delta}\rho\,\,\,,
\label{manu19}
\end{eqnarray} 
where $f_{1R_0R_0}={\rm d}^2f_1(R_0)/{\rm d}R_0^2$, $\nabla^\alpha\nabla_\alpha$ and $\nabla$ holds for the covariant derivative with respect to the unperturbed metric (\ref{I.5}). In (\ref{manu19}), we have made use of the relation linking the trace to the enery density, $T_0=\rho_0-3p_0=(1-3c_s^2)\rho_0$, and by the way, $\hat{\delta}T=(1-3c_s^2)\hat{\delta}\rho$. Here the equations of motion at the left hand side of (\ref{manu19}) presents a set of fourth-order differential equations.  By following the same assumptions, the equation of the perturbations of the continuity equation (\ref{manu14}) can be easily obtained, which yields,
\bea
\nabla_{\mu}\hat{\delta} T^{\mu}_{\nu}+\hat{\delta}\Gamma^{\mu}_{\mu\lambda} T^{\lambda}_{0\nu}-\hat{\delta}\Gamma^{\lambda}_{\mu\nu} T^{\mu}_{0\lambda}= \frac{1}{(\kappa^2-f_{2T_0})}\left\{f_{2T_0T_0}\hat{\delta} T\nabla_{\mu}T^{\mu}_{0\nu} \right. \nn
\left. +\delta^{\mu}_{\nu}\partial_{\mu}\left(\frac{1}{2}f_{2T_0}\hat{\delta} T+p_0f_{2T_0T_0}\hat{\delta} T+\hat{\delta} p f_{2T_0}\right)+\partial_{\mu}(f_{2T_0})\hat{\delta} T^{\mu}_{\nu}+T^{\mu}_{0\nu}\partial_{\mu}(f_{2T_0T_0}\hat{\delta} T)\right\}\ . 
\label{manu20}
\eea
For functions $f_2(\tilde{T})$ constant or null, the whole right hand side of the previous equation vanishes. Consequently the perturbed conservation equations become the usual ones that are obtained both in GR and $f(R)$ theories \cite{Dombriz_PRD2008} as can be inferred from (\ref{manu14}).  
\par
For the linearised equation (\ref{manu19}), the components $(ij)$, $(00)$, $(ii)$ and $(0i)\equiv (i0)$, where $i,j=1,2,3$, $i\neq j$ in Fourier space, read respectively: 

\begin{eqnarray}
\Phi-\Psi=-\frac{f_{1R_0R_0}}{f_{1R_0}}\hat{\delta}R\,\,\,,
\label{ij}
\end{eqnarray}
\begin{eqnarray}
\Big[3\mathcal{H}\left(\Phi'+\Psi'\right)+k^2\left(\Phi+\Psi\right)+3\mathcal{H}'\Psi
-\left(3\mathcal{H}'-6\mathcal{H}^2\right)\Phi\Big]f_{1R_0}\nonumber\\
+\left(9\mathcal{H}\Phi-3\mathcal{H}\Psi+3\Psi'\right)f'_{1R_0}=a^2\Big[-\kappa^2\rho_0
+\left(1-2c_s^2-3c_s^4\right)\rho_0^2f_{2T_0T_0}+\frac{1}{2}\left(3-c_s^2\right)\rho_0f_{2T_0}\Big]\delta\,\,\,,
\label{00}
\end{eqnarray}
\begin{eqnarray}
\Big[\Phi''+\Psi''+3\mathcal{H}\left(\Phi'+\Psi'\right)+3\mathcal{H}'\Phi+
\left(\mathcal{H}'+2\mathcal{H}^2\right)\Psi\Big]f_{1R_0}+\left(3\mathcal{H}\Phi
-\mathcal{H}\Psi+3\Phi'\right)f'_{1R_0}\nonumber\\
+\left(3\Phi-\Psi\right)f''_{1R_0}=a^2\Big[\kappa^2 c_s^2\rho_0+\frac{1}{2}\left(1-3c_s^2\right)\rho_0f_{2T_0}\Big]\delta\,\,\,,
\label{ii}
\end{eqnarray} 
\begin{eqnarray}
\left(2\Phi-\Psi\right) f'_{1R_0}+\Big[\Phi'+\Psi'+\mathcal{H}\left(\Phi+\Psi\right)\Big]f_{1R_0}=-a^2\left(1+c^2_s\right)\left(\kappa^2-f_{2T_0}
\right)\rho_0v\,\,,
\label{0i}
\end{eqnarray}
with
\begin{eqnarray}
\hat{\delta} R=-\frac{2}{a^2}\Big[3\Psi''+6\left(\mathcal{H}'+\mathcal{H}^2\right)\Phi
+3\mathcal{H}\left(\Phi'+3\Psi'\right)-k^2\left(\Phi-2\Psi\right)\Big]\,\,.
\end{eqnarray}
where it is easy to notice that for $f_2(T_0)=0$ the $f(R)$ equations are recovered \cite{Dombriz_PRD2008}. Moreover,  for $f_{1}(R_0)=R_0$, the GR equations are obtained \cite{Bardeen}. 
Now, by considering (\ref{manu14}) and (\ref{manu20}) in the case of $c_s^2=0$, the energy-momentum tensor conservation renders to the following first order  equations
\begin{eqnarray}
\delta'-k^2v-3\Psi' \,&=&\,-\frac{3\mathcal{H}f_{2T_0T_0}\rho_0\delta}{(\kappa^2-f_{2T_0})^2}\left(\frac{1}{2}f_{2T_0}+\rho_0 f_{2T_0T_0} \right)
+\frac{1}{\kappa^2-f_{2T_0}}\left[
\delta'\left(\frac{1}{2}f_{2T_0}+\rho_0 f_{2T_0T_0}\right)
\right.\nonumber\\
&&\left.-3\mathcal{H}\delta\left(
\frac{5}{2}\rho_0 f_{2T_0T_0}+\rho_{0}^2 f_{2T_0T_0T_0}+\frac{1}{2}f_{2T_0}\right)
\right]
\label{conservation_1_Alvaro}
\end{eqnarray}
and
\begin{eqnarray}
\Phi+\mathcal{H}v+v'\,=\,-\frac{1}{\kappa^2-f_{2T_0}}\left(\frac{1}{2}f_{2T_0}\delta+3\mathcal{H}\rho_0  f_{2T_0T_0} v\right)
\label{conservation_2_Alvaro}
\end{eqnarray}
for the temporal and spatial components respectively.
From the previous expressions is clear that for $f_{2}(T_0)\equiv 0$, the usual conservations equations in $f(R)$ 
theories (GR in particular) are recovered (see for instance eqns. (21) and (22) in \cite{Dombriz_PRD2008}). Note that expression (\ref{eqconserv}) 
has been used in order to obtain both (\ref{conservation_1_Alvaro}) and (\ref{conservation_2_Alvaro}). After further simplifications, the last two expressions
become 
\begin{eqnarray}
\delta'-k^2v-3\Psi' \,=\,0
\label{conservation_1_Alvaro_bis}
\end{eqnarray}
and
\begin{eqnarray}
\Phi+\mathcal{H}v+v'\,=\,\frac{f_{2T_0}}{2(\kappa^2-f_{2T_0})}\left(3\mathcal{H}v - \delta\right)
\label{conservation_2_Alvaro_bis}
\end{eqnarray}
that when combined yield
\begin{eqnarray}
\delta''+\mathcal{H}\left[1-\frac{3 f_{2T_0}}{2\left(\kappa^2-f_{2T_0}\right)}\right]\delta'+k^{2}\frac{f_{2T_0}}{2(\kappa^2-f_{2T_0})}  \delta + k^2 \Phi-3\Psi^{\prime\prime}-3\mathcal{H}\left[1-\frac{3 f_{2T_0}}{2\left(\kappa^2-f_{2T_0}\right)}\right]\Psi^{\prime}=\,0
\label{Combination_continuity_eqn}
\end{eqnarray}

Hence, the complete set of equations that describes the general linear perturbations for the kind of models considered here, $f(R,T)=f_1(R)+f_2(T)$, have been obtained, which  provides 
enough information about the behavior of the perturbations within this class of theories, that can be compared with expected results from 
$\Lambda$CDM model.

\section{Evolution of Sub-Hubble modes and the Quasi-static approximation}\label{Section5}

We are interested in the possible effects on the 
density contrast evolution once the perturbations enter the Hubble radius in the matter dominated era. 
In the sub-Hubble limit, i.e., $\mathcal{H}\ll k$, and after having neglected all the time derivative for the Bardeen's potentials $\Phi$ and $\Psi$, 
the equations (\ref{ij}) and (\ref{00}) can be combined yielding
\begin{eqnarray}
\Psi\,=\,\Phi\frac{1+\frac{2 k^2}{a^2}\frac{f_{1R_0R_0}}{f_{1R_0}}}{1+\frac{4k^2}{a^2}\frac{f_{1R_0R_0}}{f_{1R_0}}}\;\;\;;\;\;\; 
\Phi\,=-\frac{1}{2k^2}\left(\frac{1+\frac{4k^2}{a^2}\frac{f_{1R_0R_0}}{f_{1R_0}}}{1+\frac{3k^2}{a^2}\frac{f_{1R_0R_0}}{f_{1R_0}}} \right)\left(\kappa^2-f_{2T_0}\right)\frac{a^2 \rho_0}{f_{1R_0}} \delta \,. 
\label{Potent}
\end{eqnarray}
In addition, the equation (\ref{Combination_continuity_eqn}) in the QS approximation yields,
\begin{eqnarray}
\delta''+\mathcal{H}\left[1-\frac{3 f_{2T}^{0}}{2\left(\kappa^2-f_{2T}^{0}\right)}\right]\delta'+k^{2}\frac{f_{2T}^{0}}{2(\kappa^2-f_{2T}^{0})}  \delta + k^2 \Phi\,=\,0
\label{Combination_continuity_eqn1}
\end{eqnarray}
Then, by using the previous result (\ref{Potent}) in the equation (\ref{Combination_continuity_eqn1}) one gets
\begin{eqnarray}
\delta''+\mathcal{H}\left[1-\frac{3 f_{2T_0}}{2\left(\kappa^2-f_{2T_0}\right)}\right]\delta' +\frac{1}{2}\left[k^2\frac{f_{2T_0}}{(\kappa^2-f_{2T_0})} -
(\kappa^2-f_{2T_0})\frac{a^2\rho_0}{f_{1R_0}}\left(
\frac{1+4 \frac{k^2}{a^2}\frac{f_{1R_0R_0}}{f_{1R_0}}}{1+3\frac{k^2}{a^2}\frac{f_{1R_0R_0}}{f_{1R_0}}}\right)\right]\delta\,=\,0
\label{QSA_approximation}
\end{eqnarray}
that can be understood as the quasi-static equation for $f(R,T)$ models of the form (\ref{f2T}). By neglecting in (\ref{QSA_approximation}) 
the terms $f_2(T_0)$, i.e., paying attention only to  $f(R)$ theories, one recovers  the usual quasi-static approximation for those theories (see for instance \cite{Qstatic_Zhang}, \cite{Starobinsky} and \cite{Qstatic_Tsujikawa})
\begin{eqnarray}
\delta^{''}+\mathcal{H}\delta^{'}-\left(\frac{1+4\frac{k^2}{a^2}
\frac{f_{1R_0R_0}}{f_{1R_0}}}{1+3\frac{k^2}{a^2}\frac{f_{1R_0R_0}}{f_{1R_0}}}\right)
\frac{\kappa^2a^2\rho_0}{2f_{1R_0}}\delta \,=\,0
\label{Qstatic_strong_equation}
\end{eqnarray}
and for GR ($f_1(R_0)=R_0$), the quasi-static equation for $\delta$ becomes the well-known $k$-independent expression
\begin{eqnarray}
\delta''+\mathcal{H}\delta' - 4\pi\text{G} \rho_0 a^2\delta\,=\,0
\label{delta_0_SubHubble}
\end{eqnarray}
Note that the effect of the $f_{2}(T_0)$ terms in (\ref{QSA_approximation}) is twofold: first, the coefficient of $\delta'$ gets an extra term that depends on the first derivative of $f_{2}(T_0)$ with 
respect to $T_0$ that in general will be time dependent. Second, the coefficient for $\delta$ is also modified by adding a $k^2$ dependence that is absent
the standard quasi-static limit both in GR and in $f(R)$ theories and modifying as well the usual coefficient already present for $f(R)$ theories by a factor $(\kappa^2-f_{2T_0})$. 
The $k^2$-presence may have extraordinary consequences since for $f(R)$ theories it is usually claimed that in the two asymptotic 
limits (i.e., either GR or $f(R)$ domination), the quasi-static equation is scale independent and only in the transient regime, differences associated to the 
scale may show up. 
For the class of $f(R,T)$ theories under study, the $k^2$ term will be always dominant
for deep Sub-Hubble modes at any time of the cosmological evolution.
\par
\vspace{0.2cm}
On the other hand, a qualitative analysis taking into account that 
$\kappa^2\approx M_{P}^{-2}\approx (10^{19}{\rm GeV})^{-2}$ 
and $f_{2T_0}\approx \rho_{critical}^{-1/2}\approx (10^{-3}{\rm eV})^{-2}$,  implies that equation (\ref{QSA_approximation}) may be simplified yielding 
\begin{eqnarray}
\delta''+\frac{5}{2}\mathcal{H}\delta' +\frac{1}{2}\left[-k^2 +
f_{2T_0}\frac{a^2\rho_0}{f_{1R_0}}
\left(\frac{1+4 \frac{k^2}{a^2}\frac{f_{1R_0R_0}}{f_{1R_0}}}{1+3\frac{k^2}{a^2}\frac{f_{1R_0R_0}}{f_{1R_0}}}\right)\right]\delta\,=\,0\,.
\label{QSA_approximation_full}
\end{eqnarray}
Furthermore, if now one is interested only in extreme sub-Hubble modes, it is clear that (\ref{QSA_approximation_full}) becomes
\begin{eqnarray}
\delta''+\frac{5}{2}\mathcal{H}\delta' +\frac{1}{2}\left[-k^2 +
\frac{4}{3}f_{2T_0}\frac{a^2\rho_0}{f_{1R_0}}
\right]\delta\,=\,0
\label{QSA_approximation_full_two}
\end{eqnarray}
that in this limit and after having considered reasonable gravitational Lagrangians, i.e., not divergent, yields
\begin{eqnarray}
\delta''+\frac{5}{2}\mathcal{H}\delta' -\frac{1}{2}k^2 \delta\,=\,0\,.
\label{QSA_approximation_full_three}
\end{eqnarray}

The last expression, as well as the intermediate results  (\ref{QSA_approximation_full}) and 
(\ref{QSA_approximation_full_two}) prove that gravitational Lagrangians depending on the trace of the energy-momentum tensor
and satisfying the usual conservation equation will exhibit a density contrast evolution that is $k$-dependent for sub-Hubble modes. In comparison with the GR result (\ref{delta_0_SubHubble}), which predicts a transfer function $(T(k)\propto |\delta_k(t=t_{today})|^2)$ independent of the scale, this fact implies that the transfer function in this class of $f(R,T)$ gravities is scale-dependent. Thus, perturbations entering the Hubble horizon would become scale dependent in $k$. Therefore, models such as (\ref{viable_model_dust}), i.e., the unique models ensuring the standard conservation equation, would be theoretically excluded as will be graphically shown in the next section.
\par
In addition, note that the equation (\ref{QSA_approximation}) exhibits a singular point at $\kappa^2-f_{2T_0}=0$.  For the Lagrangian $f_2(T_0)=\alpha T_0^{1/2}+\beta$, such singular point is easily identified. From now on, let us assume the following coupling constant,
\begin{eqnarray}
\alpha \,\equiv\, c_1\kappa^{2}(\rho_{today})^{1/2}\ ,
\label{c1_def}
\end{eqnarray}
where $c_1$ is a dimensionless constant,. This parametrization is justified in order to fix the correct dimensions for the coupling constant $\alpha$. On the other hand, by solving the continuity equation (\ref{eqconserv}) for a pressureless fluid, the evolution of the matter density yields
\be
\rho_{0}=\rho_{today}\,a^{-3}= \rho_{today} (1+z)^3\ ,
\label{Bis1}
\ee
where the usual relation $1+z=a^{-1}$ has been used. Then, the expression appearing in the denominator of some terms 
in the equation (\ref{QSA_approximation}) is given by
\be
\kappa^2-f_{2T_0}=\kappa^2\left(1-\frac{c_1}{(1+z)^{3/2}}\right)\ .
\label{Bis2}
\ee
Hence, a singularity occurs at $z_s=c_1^{2/3}-1$. Then, the avoidance of such singularity constrains the value of the free parameter $c_1$:
\begin{itemize}
\item $c_1<0$, the singular point is located at $z_s<-1$, outside of the allowed range for  the redshift, as defined above.
\item $c_1>0$, here we can distinguish between two cases:  if $0<c_1<1$, then $-1<z_s<0$, and the singularity will occur in the future, while if $c_1\geq1$, the singularity is located at $z_s\geq 0$, i.e., either 
at present or past cosmological evolution.
\end{itemize}
In order to avoid any singularity, at least for the range $z>0$, we shall assume $c_1<1$. Note also that in the neighbourhood of the singularity, the equation (\ref{QSA_approximation}) reduces to,
\begin{eqnarray}
\delta''-\mathcal{H}\frac{3 f_{2T_0}}{2\left(\kappa^2-f_{2T_0}\right)}\delta' +k^2 \frac{f_{2T_0}}{2(\kappa^2-f_{2T_0})}\delta\,=\,0
\label{Bis3L}
\end{eqnarray}
and in consequence the perturbations would behave as a damped oscillator, as is analyzed in the following section and shown in Fig. 3.

 \begin{figure}[h]
\begin{center}
\resizebox{8.6cm}{8.2cm} 
{\includegraphics{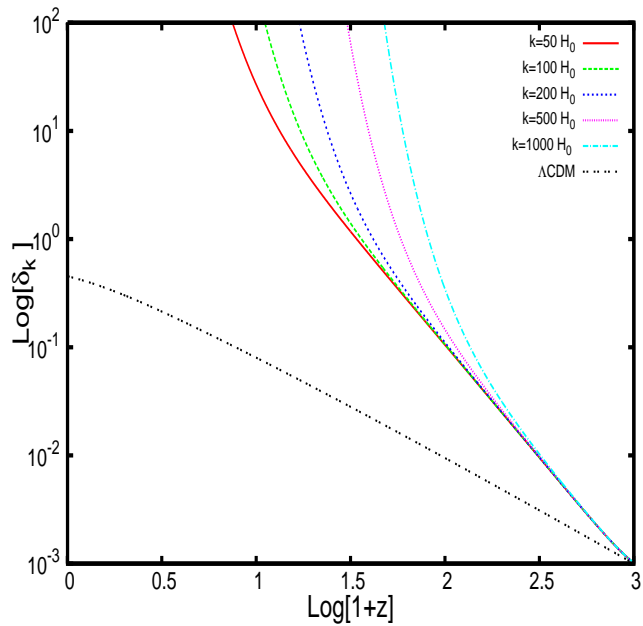}}
\caption {\footnotesize $\delta_{k}$ evolution
for $f_{A}(R,T)$ model according to the quasi-static evolution
given by (\ref{QSA_approximation_full})
and $\Lambda$CDM given by (\ref{delta_0_SubHubble}). The depicted modes are
$k= 50$, $100$, $200$, $500$ and $1000$ (in $H_0$ units). 
The plotted redshift ranged from $z=1000$ to $z=0$ (today). The value of $\Omega_m^{0}$ was fixed to 0.27 for illustrative purposes. 
It is seen how whereas the $\Lambda$CDM is $k$-independent, the 
$f_{A}(R,T)$ model evolutions diverge for all the studied modes and leave the linear region at redshifts $z\approx100$. 
For larger $k$-modes (deep Sub-Hubble modes) the divergence happens at larger redshift (earlier in the cosmological evolution).
}
\label{Fig_Model_1}
\end{center}
\end{figure}

\begin{figure}
\begin{center}
\resizebox{8.6cm}{8.2cm} 
{\includegraphics{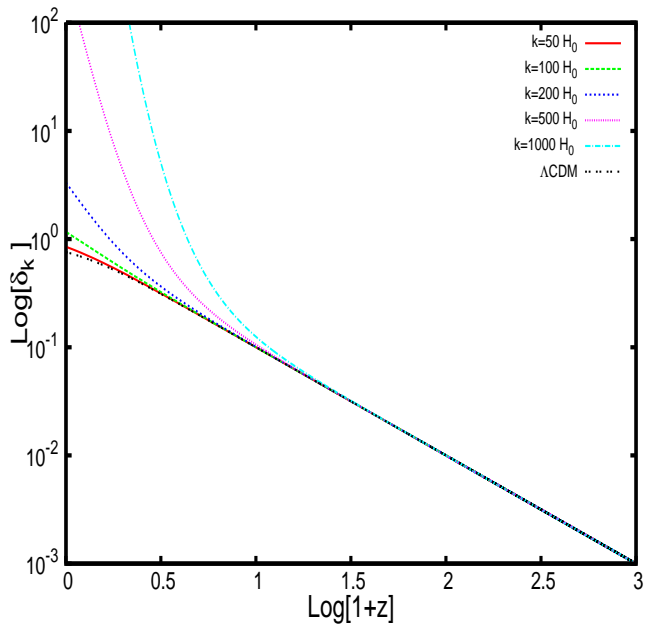}}
\caption {\footnotesize $\delta_{k}$ evolution for $f_{B}(R,T)$ model according to the quasi-static evolution given by (\ref{QSA_approximation}) and $\Lambda$CDM given by (\ref{delta_0_SubHubble}). Here we have assumed a value $c_1=-10^{-3}$. As previously, the dependence on $k$ leads to a strong growth of the matter perturbations for large values of $k$, whereas the behavior is similar to the $\Lambda$CDM model for the modes $k<200 H_0$.
}
\label{Fig_Model_2a}
\end{center}
\end{figure}

\begin{figure}
\begin{center}
\resizebox{8.6cm}{8.2cm} 
{\includegraphics{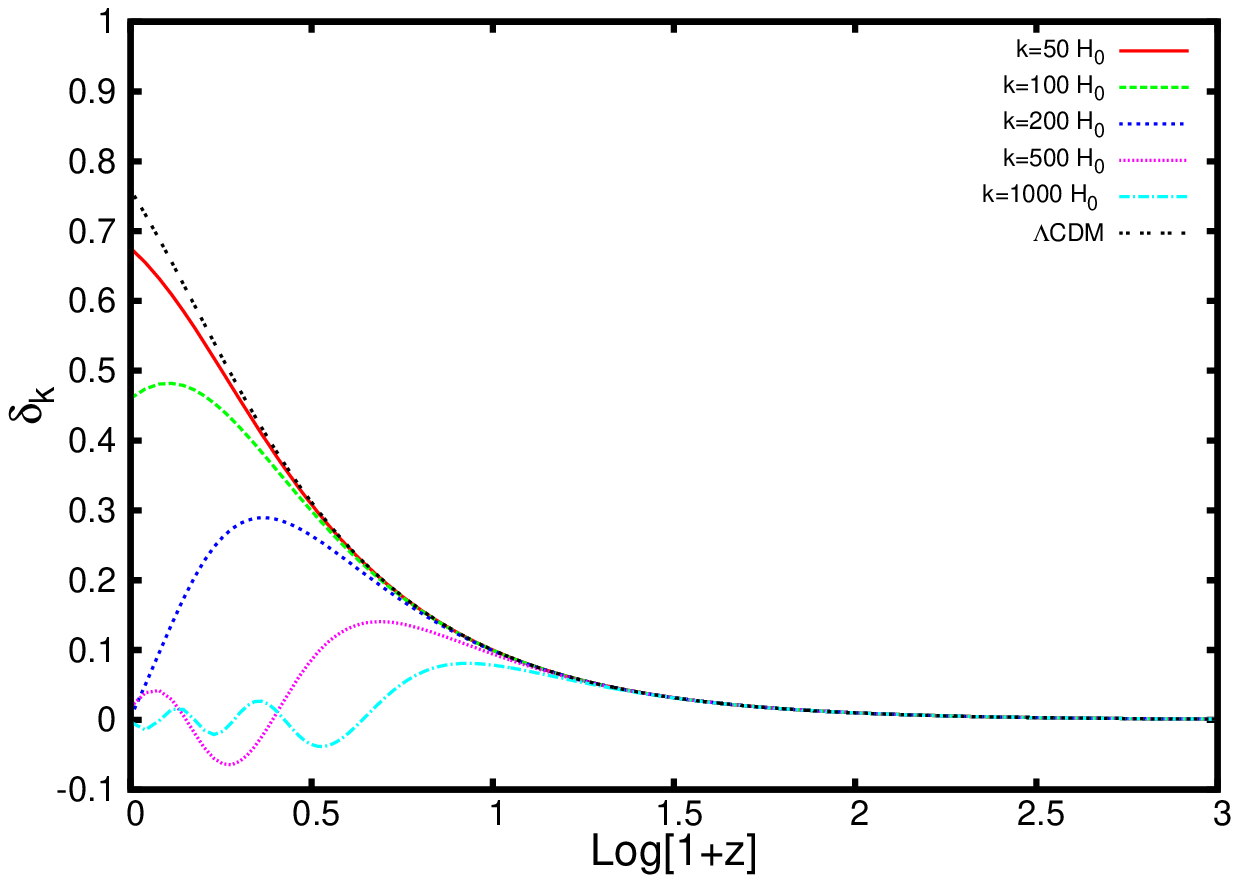}}
\caption {\footnotesize $\delta_{k}$ evolution for $f_{B}(R,T)$ model according to the quasi-static evolution given by (\ref{QSA_approximation}) and $\Lambda$CDM given by (\ref{delta_0_SubHubble}). Here we have assumed a positive value for the free parameter $c_1=10^{-3}$, which leads to an oscillating behavior of the matter perturbations, which turns out  stronger as $k$ is larger, and whose oscillations are observed for large small redshifts. The model mimics the $\Lambda$CDM model only those modes small enough $k<50 H_0$.
}
\label{Fig_Model_2b}
\end{center}
\end{figure}

\section{Numerical results}
\label{Section6}

In order to check the results obtained  in the 
previous section, we study two particular $f(R,T)$ models 
with $f_1(R_0)$ assumed to be given by 
the usual term proportional to the Ricci scalar, i.e., $f_1(R_0)=R_0$. This 
choice encapsulates a modification to GR purely originated by the function $f_2(T_0)$ introduced in Section 2 through the expression (\ref{f2T}).

\subsection{$f_{A}(R_0,T_0)=R_0+\alpha T_0^{1/2}$} 

For this function we parametrize the constant $\alpha$ according to expression (\ref{c1_def}), thus possessing the appropriate dimensions. 
In this case, one can solve the background evolution that can be rewritten as 
\begin{eqnarray}
\tilde{\mathcal{H}}^2\,=\,\Omega_{m}^{0}a^{-1}+(1-\Omega_{m}^{0})a^{1/2}
\label{constraint_eqn_Model}
 \end{eqnarray}
with $\Omega_m^0\equiv \kappa^2 \rho_{m}(\eta_{today})/3H_0^2$, the usual fractional matter density today, $H_0$ the Hubble parameter today 
and 
dimensionless conformal time defined as $\tilde{\eta}=H_0\eta$.
According to the equation (\ref{constraint_eqn_Model}), the parameter $c_1$ must accomplish
\begin{eqnarray}
c_1\,=\,-\frac{1-\Omega_{m}^{0}}{\Omega_{m}^{0}}
\label{alpha_def}
\end{eqnarray}
\par
\vspace{0.2cm}
For this model, we compare the density contrast obtained from (\ref{QSA_approximation_full}) with 
the standard $\Lambda\text{CDM}$ quasi-static approximation (\ref{delta_0_SubHubble}). The initial conditions are given at redshift 
$z = 1000$ where $\delta$ is assumed to behave as in a  matter dominated universe, i.e. $\delta_{k}(\eta)\propto a(\eta)$  with no $k$-dependence.   
In Fig. 1 we have plotted the evolution of the density contrast for several modes. One can see how the strong $k$-dependence of equation 
(\ref{QSA_approximation_full}) renders the evolution of these modes completely incompatible with the density contrast evolution
provided by the Concordance $\Lambda$CDM model and leads $\delta$ outside the linear 
order at redshift $z\approx100$.

\subsection{$f_{B}(R_0,T_0)=R_0+\alpha\, T_0^{1/2}-2\beta$}

Let us now consider the general model found in (\ref{f2T}), which also satisfies the usual continuity equation in the background but where 
the usual GR term is supplemented with a cosmological constant $-2\beta$. The first FLRW equation (\ref{FLRWf2}) yields,
\be
\tilde{\mathcal{H}}^2\,=\,\Omega_m^0 a^{-1}-c_1\Omega_m^0 a^{1/2}+c_2a^2\ ,
\label{D1}
\ee
with $c_1$ again given by (\ref{c1_def}), and $\beta\equiv 3H_0^2 c_2$  in order to provide the correct dimensions to the free constants parameters $\{\alpha,\ \beta\}$. 
By evaluating  equation (\ref{D1}) at $z=0$ (with $a(z=0)=1$), one gets the constraint,
\be
1=\Omega_m^0-c_1\Omega_m^0+c_2\quad \rightarrow \quad c_2=1-\Omega_m^0(1-c_1)\ .
\label{D2}
\ee
This expression provides a constraint on the dimensionless parameters $\{c_1, c_2\}$, where one remains arbitrary. 
%
As for the previous case, the strong dependence on $k$ in the equation (\ref{QSA_approximation}) leads to an evolution of the matter perturbations incompatible with the observations. In fact, only a very restricted limit for the free parameter $c_1$ can avoid such strong violations together with an upper limit on $k$. In Fig.~2, the case for a negative $c_1=-10^{-3}$ is considered, yielding a similar behavior as in Fig.~1. Another illustrative example of the behavior of the equation (\ref{QSA_approximation}) is shown in Fig.~3 for the value $c_1=10^{-3}$. In this case, it is straightforward to check that the equation (\ref{QSA_approximation}) turns out the damped oscillator equation for large $k$-modes, since the $k$ dependent term is positive and dominates over the other terms for small redshifts.

\section{Conclusions} \label{Section7}

In this work we have studied the evolution of matter density perturbations
in $f(R,T)$ theories of gravity. 
We have presented the required constraint to be satisfied by these theories
in order to guarantee the standard continuity equation for the energy-momentum tensor. This constraint
restricts severely the form of $f(R,T)$ models able to 
preserve  both BBN abundances and the usual behavior of both radiation and matter as cosmological fluids. 
Thus, for models of the form 
$f_1(R)+f_2(T)$ we have determined the unique $f_2(T)\propto T^{1/2}$ model able to obey the standard continuity equation.
\par
\vspace{0.2cm}

Once such viability condition was imposed in the background evolution, we have 
obtained the quasi-static approximation for these theories and shown that, for sub-Hubble 
modes the density contrast obeys a second order differential equation with strong wavenumber 
dependence. 
This fact is in contrast with well-known results for $f(R)$ fourth-order gravity theories 
and also Hilbert-Einstein action with a cosmological constant. 
\par
\vspace{0.2cm}

Then, we have compared our results with the usual quasi-static approximation in general relativity and 
shown how these two density contrasts evolve differently. 
As analyzed in the bulk of the manuscript, the quasi-static approximation equation 
may also contain a singular point forcing the matter perturbations to 
diverge along the cosmological evolution. Alternatively, the study of a positive coupling constant 
for the modified term $T^{1/2}$ led to a damped harmonic oscillator for large $k$-modes, as we illustrated in the second 
model under consideration, in particular in the case depicted in Fig.~3. 
This assumption provides 
a way to constraining the value of the coupling constant $c_1$, but does not prevent the strong deviation of the sub-Hubble 
models for this kind of models. Moreover,  the departure from the linear regime in this kind of models may happen very fast due to the explicit wavenumber 
dependence as we showed in our first studied model.  
\par
\vspace{0.2cm}
The dependence of the matter perturbations on the scale $k$ implies a great deviation with respect to those results predicted by General Relativity, giving rise to a contradiction with the  observational data provided by the main sky surveys, as for instance the Sloan Digital Sky Survey (see Ref.~\cite{other_collab}). Consequently, further analyses on these theories in the realm of cosmological perturbations evolution, including the power spectrum of the Cosmic Microwave Background (CMB) as well as the Baryonic Acoustic Oscillations (BAO) determination, would reveal the disagreement with the last observations provided by PLANCK, ruling out definitely the kind of gravitational actions studied in this investigation.

\par
\vspace{0.2cm}
Hence, our investigation concludes that models of the form $f(R,T)=f_1(R)+f_2(T)$, where the only viable $f_2(T)$ function is given by (\ref{f2T}), lead to 
 results in strong contradiction with the usually assumed behavior of the density contrast in the sub-Hubble regime, setting strong limitations for the viability of these theories and preventing this class of models to be considered as competitive candidates for dark energy.

\par
\vspace{0.2cm}

Therefore, a deep analysis of a particular theory, where the background evolution is studied alongside the cosmological perturbations, and combined with the last observations of PLANCK and the sky surveys, provides a powerful tool to discriminate the validity of alternative gravitational theories, as is the case of $f(R,T)$ gravity studied in this manuscript.

%

\begin{acknowledgments}

Authors would like to warmly thank A. L. Maroto for useful comments while the elaboration of the manuscript. 

AdlCD acknowledges financial support from  ACGC Research Fellowship University of Cape Town, MINECO (Spain) project numbers FIS2011-23000, FPA2011-27853-C02-01 and Consolider-Ingenio MULTIDARK CSD2009-00064.
MJSH thanks  CNPq-FAPES for financial support. 
MER thanks UFES for the hospitality during the elaboration of this work. 
DSG acknowledges support from a postdoctoral contract from the University of the Basque Country (UPV/EHU) under the program {\it Specialization of research staff},  and support from the research project FIS2010-15640, and also acknowledges financial support from URC (South Africa).

\end{acknowledgments}



\end{document}